\documentstyle[twocolumn]{prhep97}
\makeatletter
\let\chapter\hid@chapter
\makeatother

\input{psfig}
\begin{document}

\begin{onecolumn}
\begin{titlepage}
\begin{flushright}
  LU TP 97-35 \\
  November 1997
\end{flushright}
\vspace{25mm}
\begin{center}
  \Large
  {\bf Transverse and Longitudinal \\ Bose-Einstein Correlations in $\mbox{e}^+\mbox{e}^-$ annihilation} \\
  \normalsize
  \vspace{12mm}
  Markus Ringn\'{e}r\footnote{E-mail: markus@thep.lu.se} 
  \vspace{1ex} \\
  Department of Theoretical Physics, Lund University, \\
  S\"olvegatan 14A, S-223 62 Lund, Sweden \\
\end{center}
\vspace{20mm}

\noindent {\bf Abstract:} \\ 
We show how a difference in the
correlation length longitudinally and transversely, with respect to
the jet axis in $e^+e^-$ annihilation, arises naturally in a model for
Bose-Einstein correlations based on the Lund string model. The difference
is more apparent in genuine three-particle correlations and
they are therfore a good probe for the longitudinal stretching of
the string field. 

\vspace{2cm}
\begin{center}
(To appear in \it The Proceedings of the EPS Conference HEP97, Jerusalem, August 1997)
\end{center}

\end{titlepage}
\end{onecolumn}
\begin{twocolumn}


\authorrunning{  }
\titlerunning{  }
 


\title{Transverse and Longitudinal Bose-Einstein Correlations in $e^+ e^-$ annihilation}
\author{Markus\,Ringn\'er (markus@thep.lu.se)}
\institute{Lund University, Sweden}
\maketitle

\begin{abstract}
We show how a difference in the
correlation length longitudinally and transversely, with respect to
the jet axis in $e^+e^-$ annihilation, arises naturally in a model for
Bose-Einstein correlations based on the Lund string model. The difference
is more apparent in genuine three-particle correlations and
they are therefore a good probe for the longitudinal stretching of
the string field. 
\end{abstract}

\section{Introduction}
Bosons obey Bose-Einstein statistics, which compared to an
uncorrelated production leads to an enhancement of the production of
identical bosons at small momentum separations. The enhancement is
called the Bose-Einstein effect and it is very often parametrised in
the form
\begin{equation}
R_2(q)=1+\lambda\exp(-Q^2R^2)
\label{e:R_parametrisation}
\end{equation}
where Q is the relative four-momenta of a pair of bosons,
$Q^2=-q^2=-(p_2-p_1)^2$, and $R$ and $\lambda$ are two
phenomenological parameters. The parameter $R$ is often refered to as
the radius of the boson emitting source. It is clear that the
correlation function $R_2$ reflects the space--time region in which
the particle production occurs but both the parametrisation in
equation~\ref{e:R_parametrisation} and the interpretation of $R$ are
often given without very convincing arguments.

In reference~\cite{r:ar97} (an extension of reference~\cite{r:ah86} to
multi-boson final states) a model for Bose-Einstein correlations based
on a quantum-mechanical interpretation of the Lund string
fragmentation model \cite{r:ba83} is presented. In this work we will
investigate some features of the model to see how the correlation
lengths in the model arises and this will be used to show what the
parameter $R$ is sensitive to. We will in particular show that the
model predicts, due to the properties of string fragmentation, a
difference between the correlation length along the string and
transverse to it. In practice this means that if we introduce the
longitudinal and transverse components of the vector $q$ (defined with
respect to the thrust axis in $\mbox{e}^+\mbox{e}^-$ annihilation) we
obtain a noticable difference in the correlation distributions. This
becomes even more apparent when we go to three-particle correlations
because in this case one is even more sensitive to the longitudinal
stretching of the string field.

\section{Correlation lengths}
The starting point of our Bose-Einstein model \cite{r:ar97,r:ah86} is
an interpretation of the (non-normalised) Lund string area
fragmentation probability for an $n$-particle state (see figure~\ref{f:area})
\begin{eqnarray}
dP(p_1,p_2, \ldots,p_n) = ~~~~~~~~~~~~~~~~\\
\prod_1^n\!Ndp_j \delta(p_j^2\!-\!m_j^2)\delta(\sum p_j\!-\!P_{tot})exp(-bA) \nonumber
\label{e:lundprob}
\end{eqnarray}
in accordance with a quantum mechanical transition probability
containing the final state phase space multiplied with the 
square of a matrix element ${\cal M}$.  In reference~\cite{r:ah86} and in more
detail in reference~\cite{r:ar97} a possible matrix element is
suggested in agreement with (Schwinger) tunneling and the (Wilson) loop operators necessary to ensure gauge invariance. The matrix element is
\begin{equation}
{\cal M}=\exp(i\kappa- b/2)A
\label{e:lundM}
\end{equation}
where the area A is interpreted in coordinate space, $\kappa\simeq 1$
GeV/fm is the string constant and $b\simeq0.3$ GeV/fm is the decay
constant.

The transverse momentum properties are in the Lund model taken into
account by means of a Gaussian tunneling process. The produced
$(\mbox{q}\overline{\mbox{q}})$-pair in each vertex will in this way
obtain $\pm\bf{k}_\perp$ and the hadron stemming from the combination
of a $\overline{\mbox{q}}$ from one vertex and a $\mbox{q}$ from the
adjacent vertex obtains $\bf{p}_\perp=\bf{k}_{\perp j+1}-\bf{k}_{\perp
j}$.

In case there are two or more identical bosons the matrix element
should be symmetrised and in general we obtain the symmetrised
production amplitude
\begin{equation}
{\cal M} = \sum_{{\cal P}} {\cal M_{{\cal P}}}
\end{equation}
where the sum goes over all possible permutations of identical particles.
Taking the square we get
\begin{equation}
\label{interference}
|{\cal M}|^2\!=\!\sum_{{\cal P}} |{\cal M}_{{\cal P}}|^2 
\!\left(\!1\!+\!\!\sum_{{\cal P}^{\prime}\neq {\cal P}} 
\!\frac{2 \mbox{Re}({\cal M}_{{\cal P}}{\cal M}_{{\cal P}^{\prime}}^{\ast})}
{|{\cal M}_{{\cal P}}|^2 + |{\cal M}_{{\cal P}^{\prime}}|^2}\right)
\label{e:symM2}
\end{equation}
The MC program JETSET~\cite{r:ts94} will provide the outer sum in
equation~\ref{e:symM2} by the generation of many events but it is
evident that the model predicts a quantum mechanical interference
weight, $w_{{\cal P}}$, for each given final state characterised by
the permutation ${\cal P}$:
\begin{eqnarray}
\label{weight}
w_{{\cal P}}= 1 + \sum _{{\cal P}^{\prime}\neq {\cal P}} 
\frac{2 \mbox{Re}({\cal M}_{{\cal P}}{\cal M}_{{\cal P}^{\prime}}^{\ast})}
{|{\cal M}_{{\cal P}}|^2+|{\cal M}_{{\cal P}^{\prime}}|^2}
\end{eqnarray}
In the Lund model we note in particular for the case exhibited in
figure~\ref{f:area}, with two identical bosons denoted 1 and 2 having
a state $I$ in between, that the decay area is different if the two
identical particles are exchanged. It is evident that the interference
between the two permutation matrices will contain the area difference,
$\Delta A$, and the resulting general weight formula will be
\begin{equation} 
w_{{\cal P}} = 1+\sum_{{\cal P}^{\prime}\neq {\cal P}}\frac {\textstyle
 \cos \frac{\textstyle \Delta A}{\textstyle 2\kappa}} {\textstyle
 \cosh \left( \frac{\textstyle b\Delta A}{\textstyle 2}+
 \frac{\textstyle \Delta(\sum {\bf k}^{2}_{\perp j})}{\textstyle
 2\kappa}\right)}
\label{e:weight} 
\end{equation}
where $\Delta$ stands for the difference between the configurations
described by the permutations ${\cal P}$ and ${\cal P}^{\prime}$ and
the sum is taken over all the vertices. 
\begin{figure}
   \psfig{figure=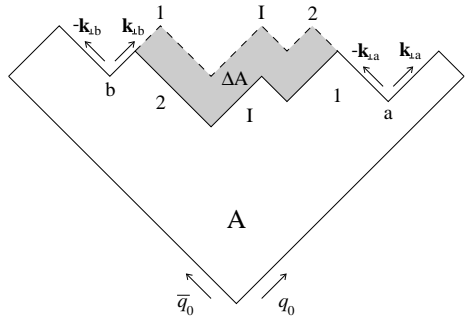,width=6.1cm}
\caption{\small \em The decay of a Lund Model string spanning the space--time
area $A$. The particles $1$ and $2$ are identical bosons and the
particle(s) produced in between them is denoted by $I$. The two ways
to produce the state are shown.}
\label{f:area}
\end{figure}
The calculation of the weight function for $n$ identical bosons
contains $n!-1$ terms and it is therefore from a computational point
of view of exponential-type. We have in reference~\cite{r:ar97}
introduced approximate methods so that it becomes of power-type
instead and we refer for details to this work.

We have seen that the transverse and longitudinal components of the
particles momenta stem from different generation mechanisms. This is
clearly manifested in the weight in equation~\ref{e:weight} where they
give different contributions. In the following we will therefore in
some detail analyse the impact of this difference on the transverse
and longitudinal correlation lengths, as implemented in the model.

In order to understand the properties of the weight in
equation~\ref{e:weight} we again consider the simple case in
figure~\ref{f:area}.  The area difference of the two configurations
depends upon the energy momentum vectors $p_1,p_2$ and $p_I$ and can
in a dimensionless and useful way be written as
\begin{equation}
\frac{\Delta A}{2\kappa} = \delta p \delta x_L 
\label{e:deltaA}
\end{equation}
where $\delta p = p_2-p_1$ and $\delta x_L = (\delta t; 0, 0,
\delta z)$ is a reasonable estimate of the space-time difference,
along the surface area, between the production points of the two
identical bosons.

To preserve the transverse momenta of the particles in the
state $(1,I,2)$ it is necessary to change the generated
${\bf k}_{\perp}$ at the two internal vertices around the state $I$
during the permutation, i.e. to change the Gaussian weights.  Also in
this case we may write a formula similar to equation~\ref{e:deltaA} for
the transverse momentum change:
\begin{eqnarray}
\frac{\Delta (\sum {\bf k}_{\perp j}^2)}{2 \kappa}= \delta {\bf p}_\perp
\delta {\bf x}_\perp
\label{e:deltakt}
\end{eqnarray}
where ${\delta {\bf p}}_\perp$ is the difference ${\bf p}_{\perp
2}-{\bf p}_{\perp 1}$ and $\delta {\bf x}_\perp = ({\bf k}_{\perp b} -
(-{\bf k}_{\perp a}))/\kappa$. The two neighbour vertices to the state
$(1,I,2)$ ($(2,I,1)$) are denoted by $a$ and $b$ and ${\bf k}_{\perp b} +{\bf
k}_{\perp a}$ corresponds to the state's transverse momentum exchange
to the outside. Therefore $\delta {\bf x}_\perp$ constitutes a
possible estimate of the transverse distance between the production
points of the pair.

For the general case when the permutation ${\cal P}^{\prime}$ is more
than a two-particle exchange there are formulas similar to
equations~\ref{e:deltaA} and \ref{e:deltakt}.

It is evident from the considerations leading to equations~\ref{e:deltaA}
and \ref{e:deltakt} that only particles with a finite longitudinal
distance and small relative energy momenta will give significant
contributions to the weights. We also note that we are in this way
describing longitudinal correlation lengths along the color fields,
inside which a given flavor combination is compensated. The
corresponding transverse correlation length describes the tunneling
(and in this model it provides a damping chaoticity).

The weight distribution we obtain is discussed in
reference~\cite{r:ar97}.  It is strongly centered around unity and we
find negligible changes in the JETSET default observables (besides the
correlation functions) by this extension of the Lund model.

\section{Results}

We have analysed two- and three-particle correlations in the
Longitudinal Centre-of-Mass System (LCMS). For each pair (triplet) of
particles the LCMS is the system in which the sum of the particles
momentum components along the jet axis is zero. In the pair analysis
we have used the kinematical variables
\begin{eqnarray}
q_\perp = \sqrt{(p_{x2}-p_{x1})^2+(p_{y2}-p_{y1})^2} \\ 
q_L = |p_{z2}-p_{z1}| \nonumber
\end{eqnarray}
and in the triplet analysis we have used
\begin{eqnarray}
q_\perp=\sqrt{q_{\perp 12}^2+q_{\perp 13}^2+q_{\perp 23}^2} \\ 
q_L=\sqrt{q_{L12}^2+q_{L13}^2+q_{L23}^2} \nonumber
\end{eqnarray}
where the jet axis is along the z-axis.

We have taken the ratio of the two-particle probability density of
pions, $\rho_2$, with and without BE weights applied as the two-particle correlation function, $R_2$
\begin{equation}
R_2(p_1,p_2)=\frac{\rho_{2w}(p_1,p_2)}{\rho_2(p_1,p_2)}
\label{e:R2}
\end{equation}
and the resulting distribution is shown in figure~\ref{f:R2}. It is
clearly seen that it is not symmetric in $q_L$ and $q_\perp$ and in
particular that the correlation length , as measured by the inverse of
the width of the correlation function, is longer in the longitudinal
than in the transverse direction. This difference remains for
reasonable changes of the width in the transverse momentum generation.
Using all the final pion pairs in the analysis results in in a small
decrease in the transverse correlation length and of course a large
decrease in the height for $q_L \simeq q_\perp \simeq 0$, while the
longitudinal correlation length is rather unaffected.
\begin{figure}
   \psfig{figure=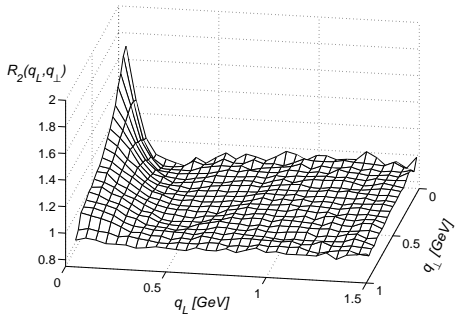,width=6.1cm}
\caption{\small \em The two-particle correlation function $R_2(q_L,q_\perp)$ for
charged pions. The sample consistes of particles which are either
initially produced or stemming from short-lived resonances.}
\label{f:R2}
\end{figure}

The total three-particle correlation function is in analogy with equation~\ref{e:R2}
\begin{equation}
R_3^{''}(p_1,p_2,p_3)=\frac{\rho_{3w}(p_1,p_2,p_3)}{\rho_3(p_1,p_2,p_3)}
\end{equation}
To get the genuine three-particle correlation function, $R_3$, the
consequences of having two-particle correlations in the model have to
be subtracted from $R_3^{''}$. To this aim we have calculated the
weight taking into account only configurations where pairs are
exchanged, $w^{'}$. In this way the three-particle correlations which
only are a consequence of lower order correlations can be defined as
\begin{equation}
R_3^{'}(p_1,p_2,p_3)=\frac{\rho_{3w^{'}}(p_1,p_2,p_3)}{\rho_3(p_1,p_2,p_3)}
\end{equation}
The genuine three-particle correlation function, $R_3$, is then given by
\begin{equation}
R_3=R_3^{''}-R_3^{'}+1
\end{equation}
This
way of getting the genuine correlations is not possible in an
experimental situation, where one has to find other ways to get a
$R_3^{'}$ reference sample. We have suggested one possible option in 
reference~\cite{r:ar97}.
\begin{figure}
   \psfig{figure=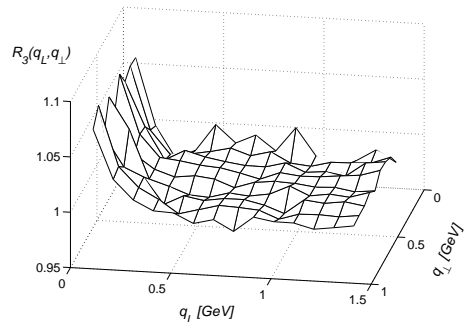,width=6.1cm}
\caption{\small \em The genuine three-particle correlation function $R_3(q_L,q_\perp)$ for all final state charged pions.}
\label{f:R3}
\end{figure}
The distribution $R_3$ is shown in figure~\ref{f:R3}.  The effect of
the higher order terms is to pull the triplets closer in longitudinal
direction while the transverse direction is rather unaffected. This
suggests that the higher order terms are more sensitive to the
longitudinal stretching of the string field.  

%

\end{twocolumn}
\end{document}